\documentclass[pra,superscriptaddress,twocolumn]{revtex4}
\usepackage{amsmath}
\usepackage{latexsym}
\usepackage{amssymb}
\usepackage{graphicx,epstopdf,subfigure,color,enumerate}
\usepackage[normalem]{ulem}
\usepackage{color}
\usepackage{mathtools}

\definecolor{Red}{rgb}{1,0,0}

\def\bra#1{\mathinner{\langle{#1}|}}
\def\ket#1{\mathinner{|{#1}\rangle}}
\def\braket#1{\mathinner{\langle{#1}\rangle}}

{\catcode`\|=\active
  \gdef\Braket#1{\begingroup
\mathcode`\|32768\let|\BraVert\left<{#1}\right>\endgroup}}
\def\BraVert{\egroup\,\mid\,\bgroup}

\def\braket#1#2{\mathinner{\langle{#1}|{#2}\rangle}}
\def\es{\mathcal{S}}

\def\ems{{\mathcal{M}_s}}
\def\emw{{\mathcal{M}_w}}
\def\eW{\mathcal{W}_g}

\def\openone{\mathbb{I}}

\def\W#1{\{#1\}_w}
\newtheorem{prop}{Property}

\begin{document}
\title{A Scheme for Performing Strong and Weak Sequential Measurements of Non-commuting Observables}
\author{Aharon Brodutch}
\email{brodutch@physics.utoronto.ca}
\address{Institute for Quantum Computing and Department of Physics and Astronomy, University of Waterloo, Waterloo, ON,  ON N2L 3G1, Canada}
	
\author{Eliahu Cohen}
\email{eliahuco@post.tau.ac.il}
\address{School of Physics and Astronomy, Tel Aviv University, Tel Aviv 6997801, Israel}
\address{H.H. Wills Physics Laboratory, University of Bristol, Tyndall Avenue, Bristol, BS8 1TL, U.K}

\begin{abstract}
Quantum systems usually travel a multitude of different paths when evolving through time from an initial to a final state. In general, the possible paths will depend on the future and past boundary conditions, as well as the system's dynamics. We present a gedanken experiment where a single system apparently follows mutually exclusive paths simultaneously, each with probability one, depending on which measurement was performed. This experiment involves the measurement of observables that do not correspond to Hermitian operators.  Our main result is a scheme for measuring these operators. The scheme is based on the erasure protocol [Phys. Rev. Lett.  116, 070404 (2016)] and allows a wide range  of sequential measurements at both the weak and strong limits. At the  weak limit  the back action of the measurement cannot be used to account for the surprising  behaviour and the resulting weak values provide a consistent yet strange account of the system's past.

\end{abstract}

\maketitle

\section{Introduction}

The joint  measurement  of sequential observables in quantum mechanics is not canonically defined. A sequential measurement of two observables at different times  can lead to incompatible results that depend on how the experiment is carried out \cite{ Aharonov1984,FHofmann2014, Diosi2015}. When the observables are \emph{compatible} (i.e., they commute)  the results of a  measurement can depend on the  \emph{context} of the measurement, leading to the well known contextuality   and non locality theorems \cite{Mermin1993,Hardy1992,Aharonov2002,Kochen1975,Guhne2010,Pusey2014}.  When the observables are incompatible, the results of measurements can be correlated in a non-trivial manner that leads to Leggett-Garg inequalities \cite{Leggett1985,Williams2008,Marcovitch2010,Goggin2011,Dresselexp2011,Suzuki2012,Groen2013,White2015,Katiyar2016} for \emph{standard} invasive measurements  and to other predictions when the measurements are less invasive \cite{Nagali2012,FHofmann2014}.

Questions regarding sequential measurements are even  more significant in systems with both past and future boundary conditions.  The observable quantities that depend on past (pre-selected) and future (post-selected) boundary conditions are not limited to Hermitian operators (or POVM elements), but rather depend on the implementation of the measurement and how it affects the measured system ({\it i.e.} the observable transition amplitudes depend on Kraus operators). The Two-State Vector Formalism (TSVF) \cite{Aharonov1964,Aharonov2007} provides a useful platform for asking questions about pre- and post-selected systems.

Aharonov, Bergman and Lebowitz (ABL) \cite{Aharonov1964} derived a formula for calculating  conditional  probabilities for the outcomes of projective measurements performed on pre- and post-selected ensembles. 
Each set of  probabilities predicted by the ABL formula is limited to a particular measurement strategy. This may lead to strange `counterfactual' predictions regarding measurements which cannot be performed simultaneously \cite{Aharonov2005}.  Aharonov, Albert and Vaidman (AAV) introduced a new way to extract information from the system with  negligible change of the measured state \cite{Aharonov1988,Duck1989,Ritchie1991}.  The result of this \emph{weak measurement} is a complex number called a \emph{weak value}.  Unlike standard observables, whose measurement result corresponds to a classical outcome, weak values can exceed the spectrum of the weakly measured operator and form an effective `weak potential'. When we weakly couple a particle to an operator $A$ describing a pre- and post-selected system, the weak value $\W{A}$ will enter the interaction Hamiltonian \cite{Aharonov2014Unusual}. Despite the `weak' method used to obtain it, each `weak value'  is related to a particular `strong' measurement strategy and can be used to understand some \emph{counterfactual} probabilities obtained using the ABL formula. This results from the following theorem \cite{Aharonov2007}: if the measured weak value of a dichotomic operator (i.e. operator having only two eigenvalues like the projections used in our gedanken experiment) is equal to one of its eigenvalues, then this value could have been found with certainty upon being strongly measured (see Property 4 below).

While strong (projective) measurements change the probability of post-selection, weak measurements introduce negligible changes of the measured state and thus hardly affect subsequent interactions with the system. For this reason, it seems that weak values of sequential observables can shed light on, and test experimentally, fundamental questions such as Leggett-Garg inequalities \cite{Leggett1985,Williams2008,Marcovitch2010,Goggin2011,Dresselexp2011,Suzuki2012,Groen2013,White2015},  which are based on multiple-times observables. Mitchison {\it et al.} have previously used weak values of sequential observables to explain strange predictions regarding measurements in a double interferometer \cite{Mitchison2007}. Their theoretical proposal has been recently further analyzed \cite{Mit2008,Aberg2009,Diosi2015} and demonstrated experimentally \cite{Pia2015}. A different approach to analyze consecutive measurements is described in \cite{Glick2009}. Here we extend the results of Mitchison et al. and relate them to the original work on sequential observables \cite{Aharonov1984}. Our main result is two-fold: a systematic analysis of the quantum system's path in terms of weak and strong sequential measurements, as well as an erasure-based scheme for implementing them. Under the right pre- and post-selection the system apparently traverses  mutually exclusive paths, each with probability $1$.

Analysis in terms of weak values requires the introduction of a new measurement scheme, first described in \cite{BCerasure}. Using the proposed scheme also helps in providing a detailed answer to the questions `what is the path of a particle?' \cite{Kocsis2011}  and  `where have the particles been?' \cite{Vaidman2013,Vaidman2014,Danan2013}. It was suggested that in-between two strong measurements, the particles have been where they left a `weak trace', {\it i.e.} a non-zero weak value.  This approach towards the past of a quantum particle has attracted much attention and  generated some  controversy \cite{Li2013,Vareply2013,Sald2014,Bart2015,Poto2015,Vareply2015a,Vareply2015b}.

While similar paradoxical behaviour has been discussed for both strong and weak measurements \cite{Aharonov1991,Vaidman2013,Vaidman2014,Danan2013}, the setup we describe is special in that the observable quantities are not Hermitian operators (see also \cite{Hu2016}).

\begin{figure}[h]
\includegraphics[width=\columnwidth]{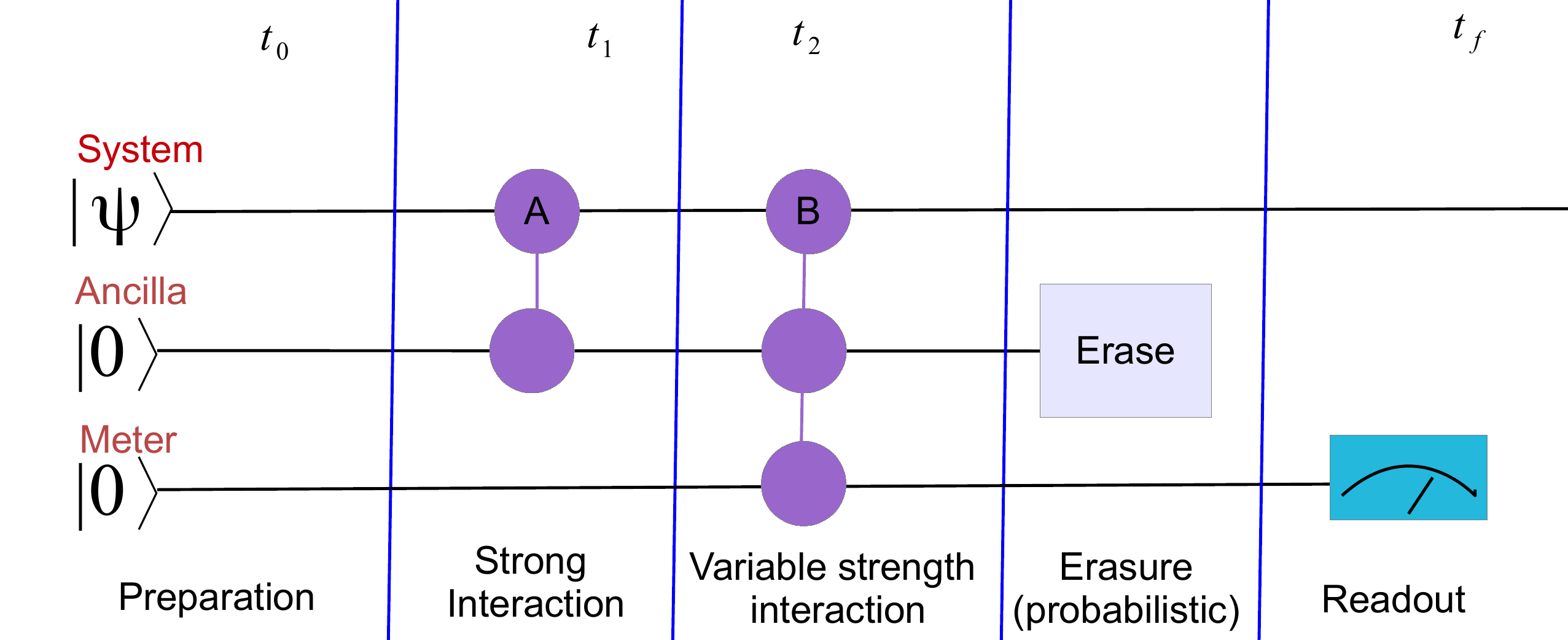}
\caption{{\bf The erasure scheme for the sequential operator $BA$}: To make a variable strength measurement of the operator $BA$ we first use an ancilla to couple strongly to $A$ at time $t_1$  and then use the meter to couple to both ancilla and the system to measure the joint observable.  The measurement is completed by erasing the result recorded in the ancilla, a probabilistic procedure.   The strength of the measurement can be selected  by changing the strength of the interaction at $t_2$. \label{fig:scheme}}
\end{figure}

Our method, based on sequential measurements of projection operators followed by quantum erasure, lets us  track the particles in several places along their way and thus provides a richer notion of their past.  To measure the sequential observable we use a technique based on the erasure scheme \cite{BCerasure} which was first used for performing non-local measurements. The setup  (see Fig. \ref{fig:scheme}) is very similar to the non-local case, first we make a strong measurement and record the result on an ancilla, we then use this result by making a measurement on both the ancilla and the system.  Finally we erase the result of the ancilla. This final step allows us to undo the backaction of the first system-ancilla interaction.  In sec \ref{sec:era} we  describe one special case in full detail. The generalization to other multi-time observables is straightforward.

The outline of the paper is as follows. In section \ref{sec1} we discuss measurements  in the TSVF. In Sec. \ref{sec2} we specialize to the case of sequential  measurements, present the Deterministic-Path Paradox and describe our main result, a scheme for performing generic sequential  measurements \cite{BCerasure}.  A reader who is only interested in the details of this scheme can skip directly to sections \ref{sec:setup} and \ref{detpath}.

\section{Measurements in the TSVF \label{sec1}}

The TSVF is a time symmetric formulation of quantum mechanics, that can be used to calculate probabilities for measurements between a pre-selection $\ket{\psi}$ and a post-selection $\bra{\phi}$. It usually helps in acquiring better understanding of a system between two projective measurements.

The following setup is relevant for  the rest of the paper: At time $t_0$ a system $\es$ is prepared in the state $\ket{\psi}$.  Later, at  $t_f$ the system undergoes a projective rank-1 measurement and is found to be in the state $\ket{\phi}$. For clarity of notation we represent this post-selected state as $\bra{\phi}$. Since quantum mechanics is time-symmetric, it is possible to think of the state $\ket{\psi}$ as evolving forward in time and on the state $\bra{\phi}$ as evolving backward in time. We assume the free evolution is the identity, so if there are no interactions at intermediate times, the state of the system is best described by the the two-state vector $\bra{\phi} \ket\psi$ \cite{Aharonov2007}.

We want to predict the results of  measurements made at the intermediate time interval $t_1$ to $t_2$, where $t_0<t_1<t_2<t_f$. We consider two types of measurements, a strong  (projective) measurement and a weak measurement. In this section we introduce the standard formalism for these two types of measurements.  In the following sections we will extend them to the case of sequential measurements using two techniques: Modular measurements \cite{Aharonov1984,Kedem2010} and the erasure method \cite{BCerasure}.

\subsection{Strong measurements}

We use the term strong measurement to refer to a measurement that gives an unambiguous result and projects $\es$ onto the corresponding subspace. Let  $A$ be a measurement operator characterized by the projectors  $A_k$ such that $A=\sum_ka_kA_k$.  Given the two-state vector $\bra{\phi} \ket\psi$, the probability that an intermediate strong measurement of  $A$ yields a result corresponding to  $A_{k}$ is given by the ABL formula \cite{Aharonov1964,Aharonov2007}:

\begin{align} \label{ABL}
Pr\left(A_{k}|\psi,\phi\right)&=
\frac{\left|\left\langle \phi|A_k|\psi\right\rangle \right|^{2}}{\sum_{j}\left|\left\langle \phi|A_j|\psi\right\rangle \right|^{2}}
\end{align}

We usually think of $A$ as an Hermitian operator and $A_k$ are projectors onto the degenerate subspaces of this operator.  However, as we will see below (Sec. \ref{detpath}), the formula can be   extended to a more general case where $A$  is a quantum channel and $A_k$ are the corresponding Kraus operators (see also \cite{Silva2014}).

\subsection{Weak measurements}

For a system described by the two-state vector $\bra{\phi} \ket\psi$,  the weak measurement of an operator $A$ produces an outcome $\W{A}=\frac{\bra\phi A \ket\psi}{\braket\phi\psi}$ called the weak value of $A$ \cite{Aharonov1988,Aharonov2007}. The standard method for implementing this weak measurement is to couple the system to a meter with momentum $P$ and position $Q$ via the von Neumann interaction Hamiltonian $H_I=f(t)AP$. This way, the shift in the pointer's position will be proportional to the weak value. The weakness (or strength) of the measurement is usually modified  by varying the strength of the interaction $g=\int f(t)dt$ and/or by changing the variance $\sigma^2$ of the initial state of the measuring device.

In general, the measurement process between $t_1$ and $t_2$ is an interaction between the system $\es$ and a meter $\mathcal{M}$, such that at the end of the process the (change in the) meter's state corresponds to the measurement outcome. While the state of the system can be arbitrary, the state of the meter is specified according to the desired properties of the measurement. To avoid confusion, we use the subscript $w$ for a meter in a weak measurement. The meter's initial state is denoted  by $\ket{0}^\emw$.

The weak measurement is a physical process parameterized by the parameter $g$. It  can be modeled as a map taking a pure product system-meter state $\rho^{\es\emw}_0=\ket{\psi}\bra{\psi}\otimes\ket{0}\bra{0}$ to a joint state $\eW(\rho^{\es\emw}_0)$, such that the following properties hold at $g\rightarrow 0$ \cite{BCerasure}.

{\prop {\bf - Non-disturbance} - The probability of post-selecting a state $\bra{\phi}$, given by $P(\phi)=tr[\bra{\phi}\eW(\rho^{\es\emw}_0)\ket{\phi}]$, is unaffected by the measurement up to terms of order $g^2$
\begin{equation}
P(\phi)=|\braket\phi\psi|^2[1-O(g^2)]
\end{equation}
}
\vspace{-10pt}
{\prop {\bf - Weak potential-} \label{prop2} After post-selection the meter state $\ket{0}^\emw$ is shifted by a value proportional to the `weak value'
\begin{equation}
\{A\}_w=\frac{\bra\phi A\ket\psi}{\braket\phi\psi}
\end{equation}

that is, the final state of the meter is $e^{-igH_{w}}\ket{0}$, where $H_w\propto \W{A}P$, thus
\begin{align}
\bra{\phi}\eW(\rho^{\es\emw}_0)\ket{\phi}\approx e^{-igH_{w}}\ket{0}\bra{0} e^{igH_{w}}+O(g^2).
\end{align}
}

The measurement must also have a strong limit at $g=1$ with the following property:

{{\prop {\bf - Strong limit-} $\mathcal{W}_1(\rho^{\es\emw}_0)$ is a strong (L\"uders) measurement.}

While most weak measurements have been realized via a coupling to continuous variables,  weak values appearing in
couplings to qubit meters have been given increasing attention recently \cite{Brun2008,Wu2009,Kedem2010,Lu2013}. In the current work we shall analyze the discrete, qubit meter case. Properties 1,2 are valid also in the case of coupling to a discrete variable, as long as the coupling is weak enough \cite{Wu2009}.  Property 3 is easily achieved when the corresponding observables are dichotomic (as in the examples that follow), however one has to be careful in more general cases.

The case of dichotomic observables is of special interest and has an interesting property in the case of weak measurements \cite{Aharonov1991}.

{\prop \label{prop4}  If $A$ is a dichotomic observable, i.e. $A=a_0A_0+a_1A_1$, where $A_i$ are projection operators, then
\begin{equation}
Pr\left(A_{k}|\psi,\phi \right)=1 \Leftrightarrow \{A\}_w=a_k
\end{equation}}
This property follows simply from property \ref{prop2} and the ABL rule, Eq. \eqref{ABL}.

\section{Sequential  measurements \label{sec2}}

As the first example of non-trivial sequential measurement Aharonov and Albert \cite{Aharonov1984} described an experiment where, by using a modular meter, it is possible to perform a correlation-type measurement of a two-time operator  $\sigma_x^{t_2}\sigma_z^{t_1}$ (that is, a measurement of $\sigma_z$ at time $t_1$ followed by a measurement of $\sigma_x$ at time $t_2$). Such a measurement can give qualitatively and quantitatively  different results than correlations between a sequence of two measurements made on two  devices, the first for $\sigma_z$ at time $t_1$ and the second for $\sigma_x$  at time $t_2$. In a more recent approach \cite{Mitchison2007}, the path of a photon through a double interferometer was shown to exhibit strange behavior when questions about its location at a given point in time were asked. These were resolved using weak values of sequential observables. The weak values were, however, related to an indirect measurement procedure. If the observable at time $t_1$ is $A_1$ and the observable at time $t_2$ is $A_2$, then it is possible to extract the sequential weak value $\W{A_2A_1}$ by performing  a weak measurements of $A_1$ using a meter $q_1$ and of $A_2$ using $q_2$.  Using an  equality  initially derived by Resch and Steinberg \cite{Resch2004a}, it can be shown that
$
\langle q_1q_2\rangle=\frac{g^2}{2}Re[\W{A_2A_1}+\W{A_1}\overline{\W{A_2}}].$ So the weak value can be extracted although the procedure  is not a weak measurement, {\it i.e.} it does not have properties 1-3 above \cite{Brodutch2009,BrodutchMsc,BCerasure,Kedem2010}.

Below we present a similar example of sequential measurements and show a scheme for performing these measurements directly. We will use this example to illustrate some of the features of weak and strong sequential measurements.   To avoid confusion we use  the notation $P(A)$ to denote the probability of outcome corresponding to the operator $A$ in a strong measurement and $\{A\}_w$ to denote the weak value of the operator $A$.

\subsection{Setup\label{sec:setup}}

Our example is based on the following scenario (see Fig. \ref{Fig1}.a): A system $\es$ is prepared in the state

\begin{align}
\ket{\psi}=\cos\theta\ket{0}+\sin\theta\ket{1}
\end{align} and post-selected in the state
\begin{align}\ket{\Phi}=\cos\phi\ket{+}+\sin\phi\ket{-},
\end{align}

where we use the standard notation $\sigma_z\ket{0}=\ket{0}$, $\sigma_z\ket1=-\ket{1}$ and $\sigma_x\ket\pm=\pm\ket\pm$.

 At the intermediate times, two projective measurements are made: First a measurement in the  $Z$  ($\{0,1\}$) basis and later a measurement in the  $X$  ($\{+,-\}$) basis. If one performs these measurements independently ({\it i.e.} each with its own meter) there are four possible results
\begin{equation}\label{results}
(0,+),~(1,+),~(0,-),~(1,-).\end{equation}

These correspond to the following four operators:
\begin{align}
A=\ket{+}\bra{+}\,\ket{0}\bra{0}=\frac{1}{\sqrt{2}}\ket{+}\bra{0}\\
B=\ket{+}\bra{+}\,\ket{1}\bra{1}=\frac{1}{\sqrt{2}}\ket{+}\bra{1}\\
C=\ket{-}\bra{-}\,\ket{0}\bra{0}=\frac{1}{\sqrt{2}}\ket{-}\bra{0}\\
D=\ket{-}\bra{-}\,\ket{1}\bra{1}=\frac{-1}{\sqrt{2}}\ket{-}\bra{1},
\end{align}
that add up to the identity $A+B+C+D=\openone$.
We also note that
\begin{align}
A+D=\frac{1}{2}(\openone -i\sigma_y)\\
B+C=\frac{1}{2}(\openone+i\sigma_y)
\end{align}
The four transition amplitudes are

\begin{align}
a=\bra{\Phi}A\ket{\psi}=\frac{\cos\theta\cos\phi}{\sqrt{2}}\\
b=\bra{\Phi}B\ket{\psi}=\frac{\sin\theta\cos\phi}{\sqrt{2}}\\
c=\bra{\Phi}C\ket{\psi}=\frac{\cos\theta\sin\phi}{\sqrt{2}}\\
d=\bra{\Phi}D\ket{\psi}=-\frac{\sin\theta\sin\phi}{\sqrt{2}}
\end{align}

\begin{figure}[h]
\includegraphics[width=0.58\columnwidth]{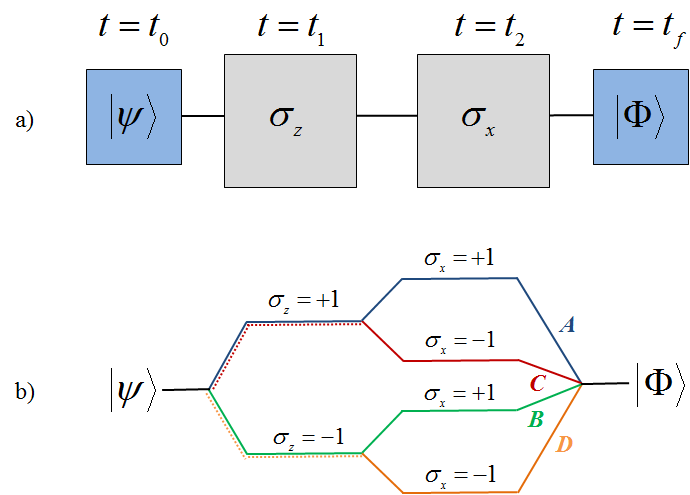}
\caption{{\bf The basic setup.} a) Two sequential measurements in the $Z$ and $X$ bases are performed on a pre- /post-selected system $\bra{\Phi} \ket\psi$. b) The four possible measurements outcomes are translated into four paths. \label{Fig1}}
\end{figure}

\subsection{A sequence of two strong measurements \label{seqstrong}}

If we use a sequence of two projective measurements ({\it i.e.} we measure in the  $Z$ basis, read out the result and then measure in the $X$ basis) we get the probabilities
\begin{subequations}\label{sequence}
\begin{align}
P(A)=\frac{a^2}{a^2+b^2+c^2+d^2}\\
P(B)=\frac{b^2}{a^2+b^2+c^2+d^2}\\
P(C)=\frac{c^2}{a^2+b^2+c^2+d^2}\\
P(D)=\frac{d^2}{a^2+b^2+c^2+d^2},
\end{align}
\end{subequations}
corresponding to the four possible paths in Fig. \ref{Fig1}.b.

This type of measurement represents the question `which path did the system follow?'. We now turn our attention to two other ways of measuring   the observables $A,B,C,D$ . The first of these (Fig. \ref{Fig2}.a)  is a measurement that can only distinguish between pairs of the results in Eq. \eqref{results}, by coupling to a single meter. In the  final  setup (Fig. \ref{Fig2}.b), each measurement  distinguishes one result from the other three, marking a distinct path from $\ket\psi$ to $\ket\phi$.  Our choice of pre- and post-selection is inspired by the three box paradox \cite{Aharonov1991} and produces a situation where we can deterministically observe the system going through two mutually exclusive paths.

\subsection{Sequential measurement using modular values \label{seqmod}}
In the spirit of \cite{Aharonov1984} we will now study  a measurement of a sequential observable $\sigma_{ZX}=\sigma_x^{t_2}\sigma_z^{t_1}$.  A result of $+1$ corresponds to both $(0,+)$ and $(1,-)$ in the $Z, X$ measurements, while a $-1$ result  corresponds  to $(0,-)$ and $(1,+)$. This is  the same as measuring the operators $M_E=A+D$  and $M_O=B+C$ respectively (see Fig. \ref{Fig2}.a).

\begin{figure}[h]
\includegraphics[width=0.58\columnwidth]{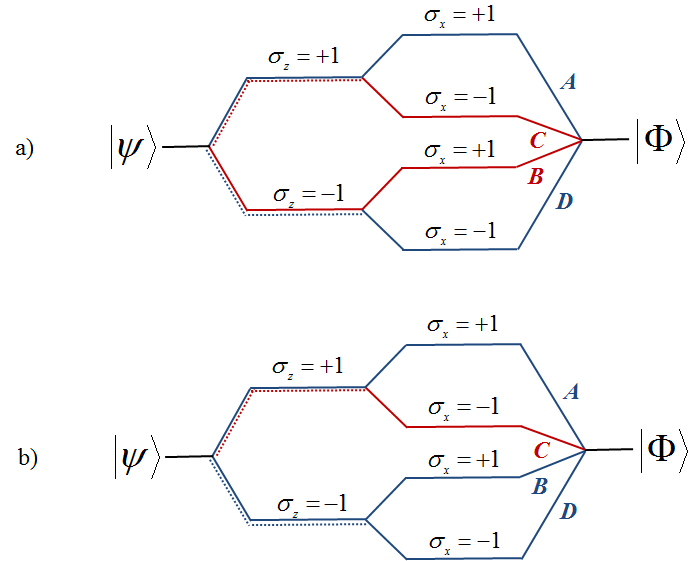}
\caption{{\bf Different measurement schemes.} a) Sequential measurement of the modular values $M_E=A+D$ and $M_O=B+C$. b) Measurement $\mathbb{C}$  of path $C$. \label{Fig2}}
\end{figure}

\subsubsection{The probabilities for a strong measurement}
The probabilities for a strong measurement are
\begin{subequations}\label{twooutcome}
\begin{align}
P(M_E)=\frac{(a+d)^2}{(a+d)^2+(b+c)^2}\\
P(M_O)=\frac{(b+c)^2}{(a+d)^2+(b+c)^2}
\end{align}
\end{subequations}

The difference between this measurement and a scheme using two separate measurements is apparent when we consider deterministic outcomes. For Eq. \eqref{sequence} we need to set $\theta=0,\pi/2$ and $\phi=0,\pi/2$. In contrast, for the modular measurement,  we need to set  either $\theta=-\phi$  to obtain  $b+c=0$  or  $\theta=\pi/2-\phi$  to obtain  $a+d=0$.

\subsubsection{The corresponding weak values}

To calculate the weak values we can use the fact that weak values are additive, hence
\begin{align}
\W{M_E}=\W{A}+\W{D}=\frac{a+d}{a+b+c+d}\\
\W{M_O}=\W{B}+\W{C}=\frac{b+c}{a+b+c+d}\\
\W{\sigma_{XZ}}=\W{M_E-M_O}=\frac{a+d-b-c}{a+b+c+d}
\end{align}

We can now see the difference between the four outcome measurement of Eq. \eqref{sequence} and the two outcome measurement of Eq. \eqref{twooutcome}. A deterministic result in the latter can be obtained in cases where either the weak values for the  individual operators are $0$ ({\it i.e.}  $\W{A}=\W{D}=0$  or $\W{B}=\W{C}=0$)  or in cases where one of the weak values is negative and balances out the other ({\it i.e.} $\W{A}=-\W{D}$ or $\W{B}=-\W{C}$).

To experimentally measure the weak and strong values for this setup it is possible to use the modular values method of Kedem and Vaidman \cite{Kedem2010}, or the erasure method \cite{BCerasure} (see Sec. \ref{sec:era} below).

\subsection{The Deterministic Path paradox\label{detpath}}

In the second example the measurement device is set in such a way (see Fig. \ref{fig:scheme} above and Sec. \ref{sec:era} below) that it clicks only for one of the four possible outcomes, say $C$ (see Fig. \ref{Fig2}.b). In this case we have four different measurement settings: $\mathbb{A}$, where a result of $1$ ( or `click') corresponds to path $A$ while a  $0$ (or `no-click') corresponds to not going through path $A$  and likewise for ${\mathbb{B}},{\mathbb{C}}$ and ${\mathbb{D}}$. Using the ABL rule we arrive at the following probabilities for a `click' in each experiment.
\begin{subequations}\label{sequential}
\begin{align}
P_{\mathbb{A}}(A)=\frac{a^2}{a^2+(b+c+d)^2}\\
P_{\mathbb{B}}(B)=\frac{b^2}{b^2+(a+c+d)^2}\\
P_{\mathbb{C}}(C)=\frac{c^2}{c^2+(a+b+d)^2}\\
P_{\mathbb{D}}(D)=\frac{d^2}{d^2+(a+b+c)^2}
\end{align}
\end{subequations}
These probabilities are very different from the sequence of individual  measurements described in Eq. \eqref{sequence}.

\subsubsection{Derivation of the `paradox'}
We want  to find  $\theta$ and $\phi$  such that $B$ and $C$ click with certainty, if the paths are projectively measured (i.e with measurements $\mathbb{B}$ and $\mathbb{C}$, respectively). Using Eq.  \eqref{sequential} the conditions for $P_{\mathbb{B}}(B)=P_{\mathbb{C}}(C)=1$ are

\begin{align}
a+c+d=0,\\
a+b+d=0,\nonumber\\
b\ne0, \;c\ne0 \nonumber
\end{align}

Subtracting these we get $c=b$,  which translates to $\cot(\theta)=\cot(\phi)$. Using this, and dividing the first equation by $\sin\theta\sin\phi \neq 0$, we get
\begin{equation}\cot^2\phi+\cot\phi-1=0\label{cond1}\end{equation}
and the solution is
\begin{equation}\label{sol}
\cot\theta=\cot\phi=\frac{-1\pm\sqrt{5}}{2}
\end{equation}

The apparent paradox can be described as:

{\it 1. The paths $B$ and $C$ are mutually exclusive.

2. The system traveled through path $B$ with certainty.

3. The system traveled through path $C$ with certainty.}

However,  statements 2 and 3 are counterfactual, we cannot ascertain them simultaneously. Or can we?

\subsubsection{The weak value of a path}
As in the three box paradox \cite{Aharonov1991}, weak measurements allow us to make sense of counterfactual statements. While the measurements  $\mathbb{B}$ and $\mathbb{C}$ cannot be performed simultaneously, their weak counterparts can.  The results are consistent with the apparent `paradox'. For general $\theta$ and $\phi$  the weak values are:
\begin{subequations}
\begin{align}
\W{A}=\frac{a}{a+b+c+d}\\
\W{B}=\frac{b}{a+b+c+d}\\
\W{C}=\frac{c}{a+b+c+d}\\
\W{D}=\frac{d}{a+b+c+d}
\end{align}
\end{subequations}

For the deterministic case, using the solution \eqref{sol}, we have
\begin{subequations}
\begin{align}
\{A\}_w&=\frac{a}{b}=\cot\theta=\frac{-1\pm\sqrt{5}}{2}\\
\{B\}_w&=\frac{b}{b}=1\\
\{C\}_w&=\frac{c}{b}=1\\
\{D\}_w&=\frac{d}{b}=-\tan\theta=\frac{2}{1\mp\sqrt{5}}
\end{align}
\end{subequations}

Hence $\W{B}=\W{C}=1$ and  $\W{A}+\W{D}=-1$. As expected  the total number of particles traversing the four paths is $1$.  Moreover,  the particle traversed both paths $B$ and $C$ with certainty.  These results  could have been predicted using property \ref{prop4}.  Furthermore, we could have used this fact to arrive at Eq. \eqref{cond1}.

This result raises a question regrading the `past of a quantum particle' \cite{Danan2013}. If the particle is understood to be at every place it left a `weak trace', then we can conclude it traveled through all four paths. However, if we understand the weak value as an effective weak potential \cite{Aharonov2014Unusual} then a weak coupling to the particle in $B$ and $C$ will result in the effective $H_I=+1$,  while coupling to the particle in $A/D$ will be negative.

This result accords well with the past results of the three box paradox \cite{Aharonov1991,Resch2004,George2013}, the negative pressure paradox \cite{Aharonov2007}, the Cheshire cat \cite{Aharonov2013,Denkmayr2014}, Hardy's paradox \cite{Hardy1992, Aharonov2002,Lundeen2009}  and others, where weak measurements reveal a curious behavior of pre- and post-selected systems.

\subsubsection{The erasure method for measuring a distinct path \label{sec:era}}

Weak values  provide an interesting perspective for the situation described above. However, their physical meaning is lost, if there is no corresponding weak measurement. To  perform the desired sequential weak measurement we use the \emph{erasure method} introduced in \cite{BCerasure}. The erasure method uses two basic principles.

1) It is possible to undo (erase) the effect of a von Neumann coupling by making an appropriate measurement on the meter.

2) It is possible to make a von Neumann measurement at any strength using the following procedure
\begin{itemize}
\item  Coupling $\es$ to  a meter $\ems$ with the standard (strong) von Neumann coupling.
\item  Coupling $\ems$ to a second meter $\emw$ using a  von Neumann coupling of the desired strength.
\item Erasing the result on $\ems$.
\end{itemize}

To show how this is done  in our example  (see also Fig. \ref{fig:scheme}) we  define three  types of unitary  operators $CNOT$ , $C_{ij}R(g)$ and $R_{ij}^{kl}(g)$ as follows:
\begin{subequations}
\begin{align}
CNOT \ket{0}\ket{\xi}&=\ket{0}\ket{\xi}\\
CNOT \ket{1}\ket{\xi}&=\ket{1}\sigma_x\ket{\xi}\nonumber
\end{align}

\begin{equation}
R(g)=e^{ig\sigma_x}
\end{equation}

\begin{align}
&C_{ij}R(g)\ket{i}\ket{j}\ket{\psi}=\ket{i}\ket{j}e^{ig\sigma_x}\ket{\psi}\\
&C_{ij}R(g)\ket{l}\ket{m}\ket{\psi}=\ket{l}\ket{m}\ket{\psi}& \forall  (i,j)\ne(l,m)\nonumber
\end{align}

\begin{align}
R_{ij}^{kl}(g)=\;& R(g)\;& if \; (i,j)=(k,l)\\
R_{ij}^{kl}(g)=&\openone\; & otherwise,\nonumber
\end{align}
\end{subequations}

where the index  $i$ takes the values $+, -$ and $j$ takes the values  $0,1$.
The measurement procedure requires an ancilla and a meter. It works in the following way:  First we perform a $CNOT$ on the system and  ancilla to measure the system in the $Z$ basis. Next we perform $C_{ij}R(g)$  with $(i,j)$ set to be the inverted sequential operator we want, for example $(i,j)=(+,0)$ for the  measurement $\mathbb{A}$ of $A$. This rotates the meter  by $g$ around  $\sigma_x$ when the system follows the measured path. Finally we erase the first measurement by post-selecting the  ancilla in the $\ket{+}$ state. Failing this final step is the same as a unitary operation on the initial state. It is possible to undo this unitary, but the cost is a change of the measurement operator.

We will follow the circuit for an arbitrary pre- and post-selection. $\ket{\psi}=\alpha\ket{0}+\beta\ket{1}$,   $\ket{\phi}=\gamma\ket{+}+\delta\ket{-}$

\begin{enumerate}
\item Initial state: 1-System; 2-Ancilla; 3-Meter
\begin{align}
\ket{\Psi_1}=[\alpha\ket{0}+\beta\ket{1}]\ket{0}\ket{0}
\end{align}

\item CNOT on 1,2

\begin{align}
\ket{\Psi_2}&=[\alpha\ket{0}\ket{0}+\beta\ket{1}\ket{1}]\ket{0}\\&\nonumber=\frac{1}{\sqrt{2}}[\alpha\ket{+}\ket{0}+\alpha\ket{-}\ket{0}+\beta\ket{+}\ket{1}-\beta\ket{-}\ket{1}]\ket{0}
\end{align}\label{after2}

\item $C_{ij}R(g)\ket{\Psi_2}$
\begin{align}\nonumber
\ket{\Psi_3}=&\frac{1}{\sqrt{2}}[\alpha\ket{+}\ket{0}R_{ij}^{+0}(g)\ket{0}+\alpha\ket{-}\ket{0}R_{ij}^{-0}(g)\ket{0}\\&+\beta\ket{+}\ket{1}R_{ij}^{+1}(g)\ket{0}-\beta\ket{-}\ket{1}R_{ij}^{-1}(g)\ket{0}]
\end{align}

\item Erasure: We measure $\sigma_x$ on 2 and discard this subsystem. If the result is $\ket{+}$  we (`succeed' and) get the unnormalized  state

\begin{align}\label{seqsuc}
\ket{\Psi_4^s}=&\alpha\ket{+}R_{ij}^{+0}(g)\ket{0}+\alpha\ket{-}R_{ij}^{-0}(g)\ket{0}\\&\nonumber+\beta\ket{+}R_{ij}^{+1}(g)\ket{0}-\beta\ket{-}R_{ij}^{-1}(g)\ket{0}
\end{align}

If the result is $\ket{-}$ we (`fail' and) get the unnormalized state
\begin{align}\label{seqfail}
\ket{\Psi_4^f}=&\alpha\ket{+}R_{ij}^{+0}(g)\ket{0}+\alpha\ket{-}R_{ij}^{-0}(g)\ket{0}\\&\nonumber-\beta\ket{+}R_{ij}^{+1}(g)\ket{0}+\beta\ket{-}R_{ij}^{-1}(g)\ket{0}
\end{align}

We will continue the derivation for successful erasure (Eq. \ref{seqsuc}) and only later return to the failed case.

\item Post-selection in $\ket{\phi}=\gamma\ket{+}+\delta\ket{-}$
\begin{align}
\ket{\Psi_5^s}=&\alpha\gamma R_{ij}^{+0}(g)\ket{0}+\alpha\delta R_{ij}^{-0}(g)\ket{0}\\&\nonumber+\beta\gamma R_{ij}^{+1}(g)\ket{0}-\beta\delta R_{ij}^{-1}(g)\ket{0}
\end{align}

We now have four cases corresponding to the different choices of $i,j$

\begin{enumerate}[(A)]
\item $(i,j)=(+,0)$
\begin{align}
\alpha\gamma R(g)\ket{0}+\alpha\delta\ket{0}+\beta\gamma\ket{0}-\beta\delta\ket{0}
\end{align}
\item $(i,j)=(+,1)$
\begin{align}
\alpha\gamma\ket{0}+ \alpha\delta\ket{0}+\beta\gamma R(g)\ket{0}-\beta\delta\ket{0}
\end{align}
\item $(i,j)=(-,0)$
\begin{align}
\alpha\gamma\ket{0}+\alpha\delta R(g)\ket{0}+\beta\gamma\ket{0}-\beta\delta\ket{0}
\end{align}
\item $(i,j)=(-,1)$
\begin{align}
\alpha\gamma \ket{0}+\alpha\delta\ket{0}+\beta\gamma\ket{0}-\beta\delta R(g)\ket{0}
\end{align}

\end{enumerate}

\end{enumerate}

Setting $\alpha=\gamma=\cos\theta$ , $\beta=\delta=\sin\theta$, and writing $R(g)\ket{0}=\ket{g}$, we have

\begin{enumerate}[(A)]
\item $(i,j)=(+,0)$
\begin{align}
\cos^2\theta\ket{g}+[2\cos\theta\sin\theta-\sin^2\theta]\ket{0}
\end{align}
\item $(i,j)=(+,1)$
\begin{align}
[\cos^2\theta+ \cos\theta\sin\theta-\sin^2\theta]\ket{0}+ \cos\theta\sin\theta\ket{g}
\end{align}
\item $(i,j)=(-,0)$
\begin{align}
[\cos^2\theta+ \cos\theta\sin\theta-\sin^2\theta]\ket{0}+ \cos\theta\sin\theta\ket{g}
\end{align}
\item $(i,j)=(-,1)$
\begin{align}
[2\cos\theta\sin\theta+\cos^2\theta]\ket{0}-\sin^2\theta \ket{g}.
\end{align}

\end{enumerate}

When $\cos^2\theta+ \cos\theta\sin\theta-\sin^2\theta=0$ the pointer in cases $B$ and $C$ will point at $g$ as expected. Note: that this is true for any value of $g$,  which is consistent with property \ref{prop4}.

Returning to the erasure step, failure would mean the post-erasure state is proportional to Eq. \eqref{seqfail}. This state corresponds to Eq. \eqref{seqsuc} with the change $\beta\rightarrow-\beta$  which can be interpreted as a $\sigma_z$ operation on the initial state.

 We can `undo' this operation by applying $\sigma_z$ to the system to get
\begin{align}
\alpha\ket{-}R_{ij}^{+0}(g)\ket{0}+\alpha\ket{+}R_{ij}^{-0}(g)\ket{0}\\\nonumber-\beta\ket{-}R_{ij}^{+1}(g)\ket{0}+\beta\ket{+}R_{ij}^{-1}(g)\ket{0}
\end{align}

Comparing with Eq. \eqref{seqsuc} we have the same state up to  a change in the measurement operator, switching  $\mathbb A \leftrightarrow \mathbb C$ and $ \mathbb B \leftrightarrow \mathbb D$.  For a weak measurement we can apply $\mathbb A,\mathbb B,\mathbb C$ and $\mathbb D$ simultaneously so that the change in measurement operator can be corrected in post-processing.

Similarly, the above method can be used for performing non-local weak measurements of the peculiar weak values discussed in \cite{Aharonov2012Pec}.

Finally, it is possible to think of other interesting  measurement operators that can create stronger versions of the paradox above. However,  it is unclear if we can assign physical meaning to those operators using  the \emph{erasure scheme} or other more complex schemes.


\section{Conclusions and outlook}

We presented an extension of the erasure scheme \cite{BCerasure}  that can be used to perform special types of sequential measurements. To demonstrate the  potential of this  scheme we described three  gedanken experiments that showcase the difference between a sequence of measurements and sequential measurements. In the first we ask  `which path did the system go through?', in the second we ask  `did the system go through $A/D$ or through $B/C$?'. Finally, for each path we ask  `did the system go through this path?'.  The answers to these three questions do not agree, and moreover, the answer to the last question suggests that the system exhibits surprising behavior by deterministically traversing two routes at the same time. It is possible to attribute the  strange behavior to the fact that the questions are mutually exclusive, however the results are consistent even when weak measurements are used. Our scheme shows  that these theoretical predictions could be can be tested experimentally.

Optical implementations of weak measurement experiments have become common in the last decade \cite{Goggin2011, Danan2013, Resch2004, Lundeen2009, Boyd2013}. It is convenient to think of an optical version of the paradox using a double interferometer as in \cite{Mitchison2007}. The advantage of this setup is that the path described by the $Z,X$ measurements corresponds to an actual path along the arms of the interferometer.  However, from an experimental point of view an optical test of this experiment may be hard to realize due to the large number of qubits  required ({\it i.e.} 3-5 qubits \footnote{Although the experiment involves 3 qubits, 2 additional ancilla qubits are required in some platforms such as liquid state  NMR  to perform the post-selection}).    The  mathematical description of our protocol in terms of Pauli operators (rather than optical paths) was deliberately chosen with the view that these experiments are more likely to be performed in other platforms such as liquid state NMR \cite{Lu2013} or superconducting qubits (where the sequential weak measurements on independent meters are performed routinely)  \cite{Groen2013,White2015,Dresselqu2014}. References \cite{Groen2013,White2015} experimentally demonstrate a qubit meter in a superconducting system with explicitly variable strength, with the latter experiment being realized on a chip with 9 independently controllable qubits, thus possibly allowing the proposed circuit to be implemented in a similar system.

The experiment described here demonstrates only one potential use of the erasure scheme in the context of sequential or non-Hermitian observables.  The study of these types of observables might have been overlooked in the past due to the lack of a method to make sequential measurements and partly due to the ambiguity in their definition as described in the three different experiments above. The erasure scheme opens up the possibility for making a large class of such sequential measurements, as well as the possibility for further exploration of their significance. They may lead to two complementary research directions, one related to the foundations of quantum mechanics and one related to development of practical methods in quantum information processing and quantum control.

On the foundational side, sequential measurements of the type described in Sec. \ref{detpath} offer a different  view of the evolution of a quantum system and provide a unique approach for probing the causal structure of quantum mechanics. They can be combined with the non-local measurements described in \cite{BCerasure} to probe  general space-time observables (e.g. in the cases of quantum steering \cite{Sch1935,Wis2007} or summoning \cite{Kent2013,Hayden2012}) in a way that has not been explored yet in much detail (see also \cite{Glick2009} for one implication of these sequential measurements). General space-time observables  are interesting at both the weak and strong regimes. It is, however, unclear if the erasure scheme can be somehow modified to perform a continuous sequential measurement of this type which could provide answers to more general questions.

Regarding quantum control, we have already described how the non-local version of the erasure scheme can be used for performing a variable strength Toffoli gate \cite{BCerasure}.  Examples of this kind are easy to find when discussing non-local interactions since  they are very natural in quantum protocols. The role of sequential observables in quantum information processing and quantum control methods have not been explored much, but we believe that our experimentally feasible method is capable of opening up new opportunities in this direction.

\vspace{15pt}

\begin{acknowledgements}
We would like to thank Yakir Aharonov, Raymond Laflamme, Marco Piani and Lev Vaidman for their valuable input on this work. AB is supported by NSERC, Industry Canada and CIFAR. EC was supported in part by the Israel Science Foundation Grant No. 1311/14 and by ERC AdG NLST. AB is now at the University of Toronto's Center for Quantum Information and Quantum Control.
\end{acknowledgements}




\end{document}